# Generation of polarization squeezed light with an optical parametric amplifier at 795 nm

Yashuai Han[1, 2], Xin Wen[1, 2], Jinyu Liu[1, 2], Jun He[1, 2, 3], and Junmin Wang[1, 2, 3, *]

1). State Key Laboratory of Quantum Optics and Quantum Optics Devices (Shanxi University),
2). Institute of Opto-Electronics, Shanxi University,
3). Collaborative Innovation Center of Extreme Optics (Shanxi University),
No.92 Wu Cheng Road, Tai Yuan 030006, Shan Xi Province, People's Republic of China

*Corresponding author, email: wwjjmm@sxu.edu.cn

**Abstract:** We report the experimental demonstration of polarization squeezed beam at 795 nm by combining a quadrature squeezed beam with a coherent beam. The quadrature squeezed beam is generated by a degenerate optical parametric amplifier based on a PPKTP crystal. Stokes parameter squeezing of -3.8 dB and anti squeezing of +5.0 dB is observed. This polarization squeezed beam resonant to rubidium $D_1$ line has potential applications in quantum information networks and precise measurement beyond the shot noise limit.

## 1. Introduction

As an important nonclassical light field, the squeezed light has been widely used in the fields of quantum teleportation [1-3], quantum dense coding [4-6], quantum storage [7-9], gravitational wave detection [10-12] *and etc*, since it is experimentally generated in 1985 [13]. In the field of squeezed light, in addition to quadrature squeezed light, photon number squeezed light and intensity difference squeezed light, the recently developed polarization squeezed light has also been widely concerned. The polarization squeezed light can be understood as a linear polarized coherent light, whose orthogonal vacuum channel is filled by a squeezed vacuum. In the field of quantum storage, the polarization squeezed light can directly interact with the spin wave of atomic ensemble. In addition, the polarization squeezed light can also improve the sensitivity of the magnetic field measurement [14]. At present, the sensitivity of magnetometer based on atomic spin effect reaches 0.16 fT/Hz$^{1/2}$ [15], which is close to the shot noise limit. To further improve the sensitivity, two fundamental quantum noises must be considered, including atomic spin projection noise and polarization noise of the light field. The polarization squeezed light can realize the sensitivity of the magnetic field measurement beyond the shot noise limit.

In 1993, Chirkin *et al.* [16] first characterized the polarization state of the two-mode field by the Stokes operators and discussed the concept of polarization squeezing for continuous variable (CV). After then, many groups carried out related experimental research. For now, there are three schemes to realize the polarization squeezed light. The first scheme is based on the Kerr nonlinearity of the fiber [17-19]. Using this scheme, up to 5.1dB of pulsed polarization squeezing at 1497 nm was observed [19]. The second scheme is based on the polarization self-rotation (PSR) effect of atomic ensemble [20-22]. Via PSR, -3 dB of CV polarization squeezing has been observed for $D_1$ transitions using $^{87}$Rb vapor [22]. The last scheme is based on the second-order nonlinear effect of nonlinear crystal [14, 23-26]. Compared with results base on PSR effect of atomic ensemble, the generated polarization squeezed light with this scheme has the advantage of good tenability and higher squeezing degree. In 2006, Wu *et al.* [26] generated -4.0 dB of polarization squeezed light at 795 nm by combining two quadrature amplitude squeezed optical beams on a polarization beam splitter (PBS) cube. This is the maximum polarization squeezing observed at this wavelength. However, this scheme needs two optical parametric amplifiers (OPA), which increases the complexity and cost of the system.

In this paper, we generated the CV polarization squeezed light resonant with the Rb $D_1$ line by combining the amplitude squeezed beam with an orthogonally polarized bright coherent beam. A 3.8 dB polarization squeezing with Stokes operator $\hat{s}_2$ is achieved. Although the squeezing degree is a little bit lower compared with Wu *et al*'s work [26], the experimental system is greatly simplified. This polarization squeezed light at Rb $D_1$ line has huge potential in precise measurement of magnetic field.

## 2. Polarization squeezing

The polarization state of a light beam can be described by four Stokes operators ($\hat{s}_0$, $\hat{s}_1$, $\hat{s}_2$, and $\hat{s}_3$) on a Poincaré sphere [16]. $\hat{s}_0$ represents the intensity of the beam, the other three

Stokes operators characterize the polarization and form a Cartesian axis system. $\hat{S}_1$, $\hat{S}_2$ and $\hat{S}_3$ stand for horizontally, linearly at 45°, and right-circularly polarized, respectively. Following the references [27] and [28], the Stokes operators can be expanded in term of the annihilation $\hat{a}$ and creation $\hat{a}^\dagger$ operators of the horizontally (H) and vertically (V) polarized modes. The results are as follows:

$$\hat{S}_0 = \hat{a}_H^\dagger \hat{a}_H + \hat{a}_V^\dagger \hat{a}_V, \quad \hat{S}_1 = \hat{a}_H^\dagger \hat{a}_H - \hat{a}_V^\dagger \hat{a}_V$$

$$\hat{S}_2 = \hat{a}_H^\dagger \hat{a}_V e^{i\theta} + \hat{a}_V^\dagger \hat{a}_H e^{-i\theta}, \quad \hat{S}_3 = i\hat{a}_V^\dagger \hat{a}_H e^{-i\theta} - i\hat{a}_H^\dagger \hat{a}_V e^{i\theta} \qquad (1)$$

Where $\theta$ is the relative phase between H and V polarized modes. If the quantum noise of these orthogonally polarized modes is uncorrelated, the variances of Stokes operators are given by

$$V_0 = V_1 = \left\langle \alpha_H^2 (\delta \hat{X}_H^+)^2 \right\rangle + \left\langle \alpha_V^2 (\delta \hat{X}_V^+)^2 \right\rangle$$

$$V_2(\theta=0) = V_3(\theta=\frac{\pi}{2}) = \alpha_V^2 \left\langle (\delta \hat{X}_H^+)^2 \right\rangle + \alpha_H^2 \left\langle (\delta \hat{X}_V^+)^2 \right\rangle$$

$$V_3(\theta=0) = V_2(\theta=\frac{\pi}{2}) = \alpha_V^2 \left\langle (\delta \hat{X}_H^-)^2 \right\rangle + \alpha_H^2 \left\langle (\delta \hat{X}_V^-)^2 \right\rangle \qquad (2)$$

Here, $\delta \hat{X}_{H,V}^+$ and $\delta \hat{X}_{H,V}^-$ are the quadrature amplitude and phase noise operators of two orthogonally polarized modes, respectively. In our experiment, the phase difference is locked to 0. In this case, we combine an s-polarized amplitude squeezed beam with a p-polarized strong coherent beam on a PBS cube. According to the equation (2), we expect the variances to be: $V_0 = V_1 = 1$, $V_2 < 1$ and $V_3 > 1$, where the quantum noise limit (QNL) is normalized to be 1. If the phase difference is locked to $\pi/2$, the variances of the stokes operators turn into: $V_0 = V_1 = 1$, $V_2 > 1$ and $V_3 < 1$. Fig. 1 shows the representations of quantum polarization states for the generated polarization squeezed beam for the two circumstances. Fig. 1(a) represents the circumstance for the phase difference of $\pi/2$, and Fig. 1(b) is the result for the phase difference of 0.

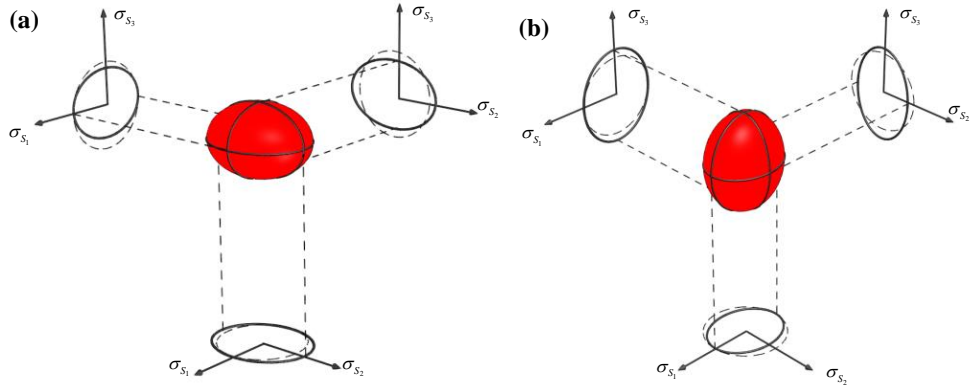

**Fig. 1** Representations of quantum polarization states for the generated polarization squeezed beam for (a) $\hat{S}_2$ is anti-squeezed and $\hat{S}_3$ is squeezed; (b) $\hat{S}_2$ is squeezed and $\hat{S}_3$ is anti-squeezed. For both (a) and (b), $\hat{S}_0$ and $\hat{S}_1$ is QNL. The red ellipsoid is the squeezed noise ball, the three ellipses are the projections at each plane. The dashed circle represents the noise of coherent state and the solid one shows the squeezing results.

## 3. Generation and detection of polarization squeezing

The schematic of experimental setup is shown in Fig. 2. The fundamental source is a continuous-wave Ti:Sapphire laser made by M squared Ltd. The wavelength of the laser is adjusted to 794.975 nm, resonant with Rb $D_1$ line. An isolator and phase-type electro-optic modulator (EOM) are used at the output of the laser. Although the follow-up enhancement cavities are bow-tie configuration, the feedback is serious and an isolator is required. The EOM is utilized to realize the Pound Drever – Hall locking of all the enhancement cavities. Then the main laser is divided and injected into three enhancement cavities, including a second-harmonic generation (SHG) cavity, an optical parametric amplifier (OPA) cavity and a mode-cleaner (MC) cavity. Two PPKTP crystals (Raicol Crystals Ltd.) with a poling period of 3.15 μm are positioned at the centre of the concave mirrors for SHG and OPA cavities. The PPKTP crystals are placed in copper-made ovens, whose temperature was actively stabilized to about 53°C for optimization. PPKTP is chosen due to a high nonlinearity together with a relatively good transmission at 397.5 nm. These three cavities are all designed in a uniform bow-tie configuration, which have lengths of about 600 mm and a distance of around 120 mm between the two concave mirrors. This gives a beam waist of ~40 μm at the central of the crystals. The transmissivity of the input coupler of SHG cavity is 10% at 795 nm. For OPA cavity, the transmissivity of the output coupler is 11.5%. The detailed parameters of these three cavities can be seen in our previous paper [29]. The MC is used to filter the spatial mode of the local beam, ensuring a high interference visibility in the homodyne detection.

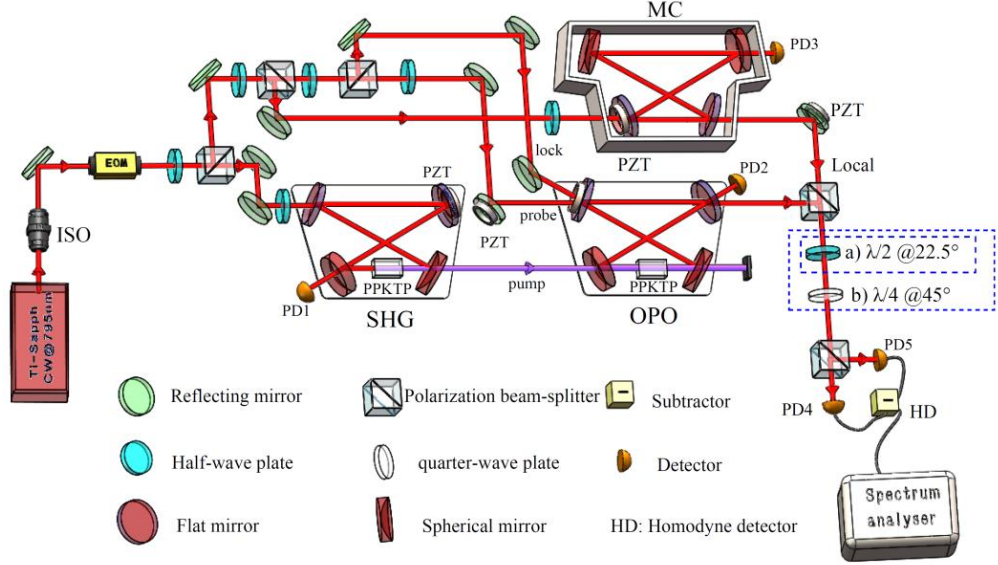

Fig. 2 Schematic of experimental setup. Optics in (a) and (b) are included to measure the variance of $\hat{S}_2$ and $\hat{S}_3$, respectively.

The infrared laser is efficiently frequency-doubled in the SHG cavity. More than 100 mW of ultra-violet (UV) laser at 397.5 nm can be obtained at the fundamental power of 190 mW. Then, the 397.5 nm UV laser is mode-matched to the OPA cavity. In our previous paper [29, 30], we mentioned that strong grey tracking effect was observed at high UV power for both SHG and OPA process. To minimize the grey tracking effect, we restrict the UV power to around 50 mW. By controlling the relative phase between the pump and probe laser, the probe laser is either amplified or de-amplified. At the UV pump power of 49 mW, an amplification factor of 3.2 and a de-amplification factor of 0.47 are measured. We lock the relative phase to de-amplification in order to generate the quadrature amplitude squeezed beam. This squeezed beam is combined with a bright coherent beam with a PBS. Then, the combined beam is directed to the polarization measurement device.

The combined beam is split on a PBS and detected with a pair of photodiodes with typical quantum efficiency of 95%. The photocurrents of two outputs are added and subtracted, which gives the values of $\hat{S}_0$ and $\hat{S}_1$, respectively. To measure $\hat{S}_2$, the polarization of the beam is rotated by 45° with a half-wave plate and the photocurrents are subtracted, shown in Fig. 2(a). To measure $\hat{S}_3$, apart from the half-wave plate rotation, a quarter-wave plate is introduced to generate a π/2 phase shift between s and p polarization, then the photocurrents are subtracted again, shown in Fig. 2(b). The variances of these stokes parameters are measured by sending the output photocurrent into a spectrum analyzer (Agilent, Model E-4405B).

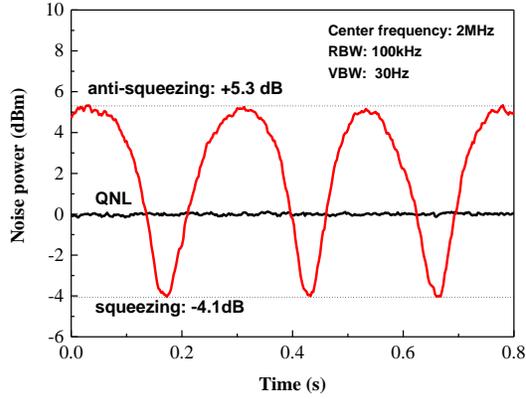

Fig. 3 Measured noise power of $\hat{S}_2$, as the phase of the local oscillator beam is scanned. The analyzing frequency is 2 MHz in the zero-span mode with RBW= 100 kHz, and VBW=30 Hz.

We first measure the squeezing traces of $\hat{S}_2$ in a scanning mode by scanning the relative phase between the squeezed vacuum and the local beam, shown in Fig. 3. The squeezing and anti-squeezing levels are -4.1 dB and +5.3 dB at the center frequencies of 2 MHz, respectively. With a quantum efficiency of 95%, escape efficiency of 96.6%, propagation efficiency of 99%, and an interference visibility of 99.7%, the expected squeezing level at 2 MHz is -5.6 dB for the UV power of 49 mW. This discrepancy may be attributed to the unexpected parametric gain and the increase of intracavity losses with the UV pump power, which is discussed in our previous paper [29]. Fig. 4 shows the measured variances of four stokes parameters in a locking mode. The OPA is locked to de-amplification and the relative phase between the generated amplitude squeezed beam and the bright coherent beam is locked to 0. Each trace is normalized to QNL and the measurements are taken at 2 MHz in the zero span mode with RBW=100 kHz and VBW=30 Hz. According to equation (2), the variance of $\hat{S}_0$ and $\hat{S}_1$ should be at the QNL. However, the measured results for $\hat{S}_0$ and $\hat{S}_1$ shown in Fig. 4(a) is slightly above the QNL, which is induced by the residual classical noise of probe laser at 2 MHz. The $\hat{S}_2$ parameter is squeezed by -3.8±0.04 dB relative to the QNL, while $\hat{S}_3$ is anti-squeezed by +5.0±0.02 dB, which is shown in Fig. 4(b) and Fig. 4(c), fitting the theoretical expectation.

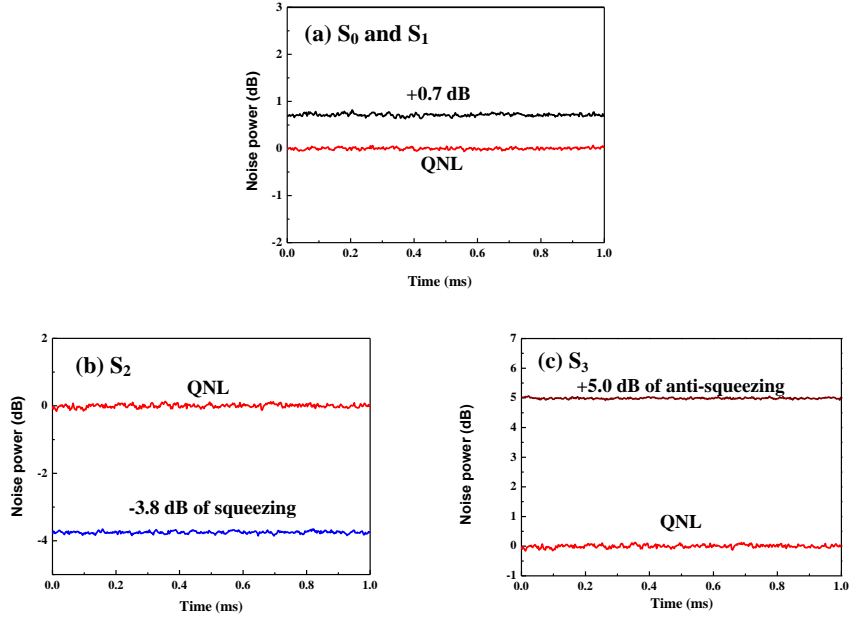

Fig. 4 Quantum noises of stokes operators of polarization squeezed beam. (a) $\hat{S}_0$ and $\hat{S}_1$; (b) $\hat{S}_2$; (c) $\hat{S}_3$. The measurements are taken at 2 MHz in the zero span mode with RBW=100 kHz and VBW=30 Hz. The data is averaged for 20 times.

## 4. Conclusion

We have demonstrated the generation of polarization squeezing resonant to the rubidium $D_1$ line using an OPA cavity with PPKTP crystal. Combining the quadrature amplitude squeezed beam with the bright coherent beam, -3.8 dB of squeezing in $\hat{S}_2$ and +5.0 dB of anti-squeezing in $\hat{s}_3$ was observed. We hope the generated polarization squeezed beam at 795 nm can be applied in the spin exchange relaxation free atomic magnetometer to realize the sensitivity of the magnetic field measurement beyond the shot noise limit.